%% file: main.tex
\documentclass[journal]{IEEEtran}
\usepackage{amsmath}
\usepackage{amsfonts}
\usepackage{algorithmic}
\usepackage{array}
\usepackage{graphicx}
\usepackage{color}
\ifCLASSOPTIONcompsoc
 \usepackage[caption=false,font=normalsize,labelfont=sf,textfont=sf]{subfig}
\else
 \usepackage[caption=false,font=footnotesize]{subfig}
\fi
\usepackage[hyphens]{url}

\usepackage{breakurl}
\usepackage[breaklinks]{hyperref}

\newcommand{\revise}[1]{#1}

\begin{document}

\title{Real-Time Prediction of the Duration of Distribution System Outages}
\author{Aaron Jaech,
        Baosen Zhang, {\it Member, IEEE},
        Mari Ostendorf, {\it Fellow, IEEE}
        and Daniel S. Kirschen, {\it Fellow, IEEE}
\thanks{The authors are with the Department of Electrical Engineering, University of Washington, Seattle, WA 98195
 e-mails: \{ajaech,zhangbao,ostendor,kirschen\}@uw.edu}
}

\maketitle

\begin{abstract}
This paper addresses the problem of predicting duration of unplanned power outages, using historical outage records to train a series of neural network predictors. The initial duration prediction is made based on environmental factors, and it is updated based on incoming field reports using natural language processing to automatically analyze the text. Experiments using 15 years of outage records show good initial results and improved performance leveraging text. Case studies show that the language processing identifies phrases that point to outage causes and repair steps.
\end{abstract}


\section{Introduction}
\label{sec:intro}
\input{intro.tex}

\section{Problem Definition}
\label{sec:setup}
\input{setup.tex}

\section{Outage Duration Predictors}
\label{sec:model}
\input{model.tex}

\section{Experiments}
\label{sec:expts}
\input{expts.tex}

\section{Analysis and Case Studies} \label{sec:analysis}
\input{analysis.tex}

\section{Conclusions}
\label{sec:conclusion}
\input{concl.tex}

\subsection*{Acknowledgments}

This work was supported by NSF Award \#1509880. The authors thank Seattle City Light for providing the outage and repair logs that made this research possible, and Ruchira Kulkarni for the initial work on processing the repair logs. The views, opinions and positions expressed by the authors are theirs alone, and do not necessarily reflect the views, opinions or positions of NSF or Seattle City Light.

\Urlmuskip=0mu plus 1mu\relax
\bibliographystyle{IEEEtran}
\bibliography{mybib}

\begin{IEEEbiography}
	[
	{\includegraphics[width=1in,height=1.25in,clip,keepaspectratio]{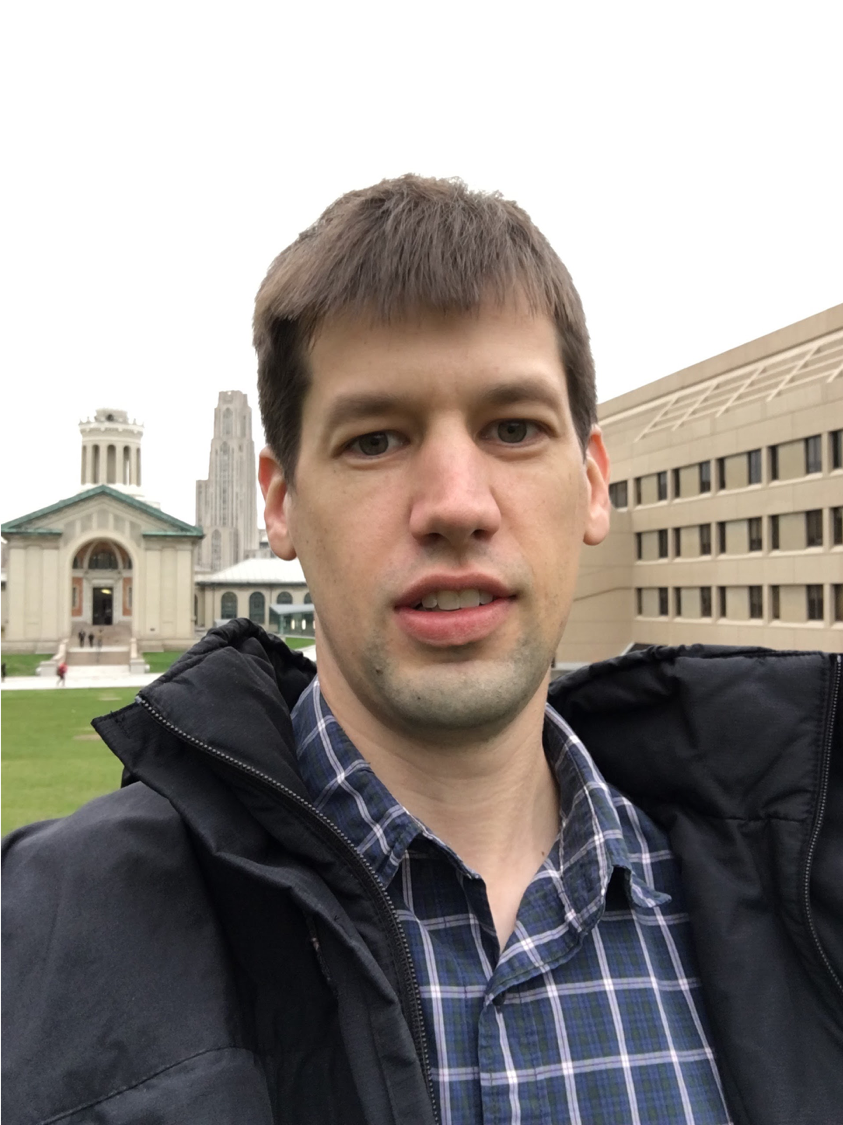}}
	]
  {Aaron Jaech} received the B.Sc. degree from Carnegie Mellon University in Mathematical Sciences and Computer Science in 2011. He completed his Ph.D. in Electrical Engineering at University of Washington in 2018 under the advisement of Prof. Mari Ostendorf. His research interests include natural language and speech processing. He is currently working as a research scientist at Facebook.
\end{IEEEbiography}

\begin{IEEEbiography}
	[
	{\includegraphics[width=1in,height=1.25in,clip,keepaspectratio]{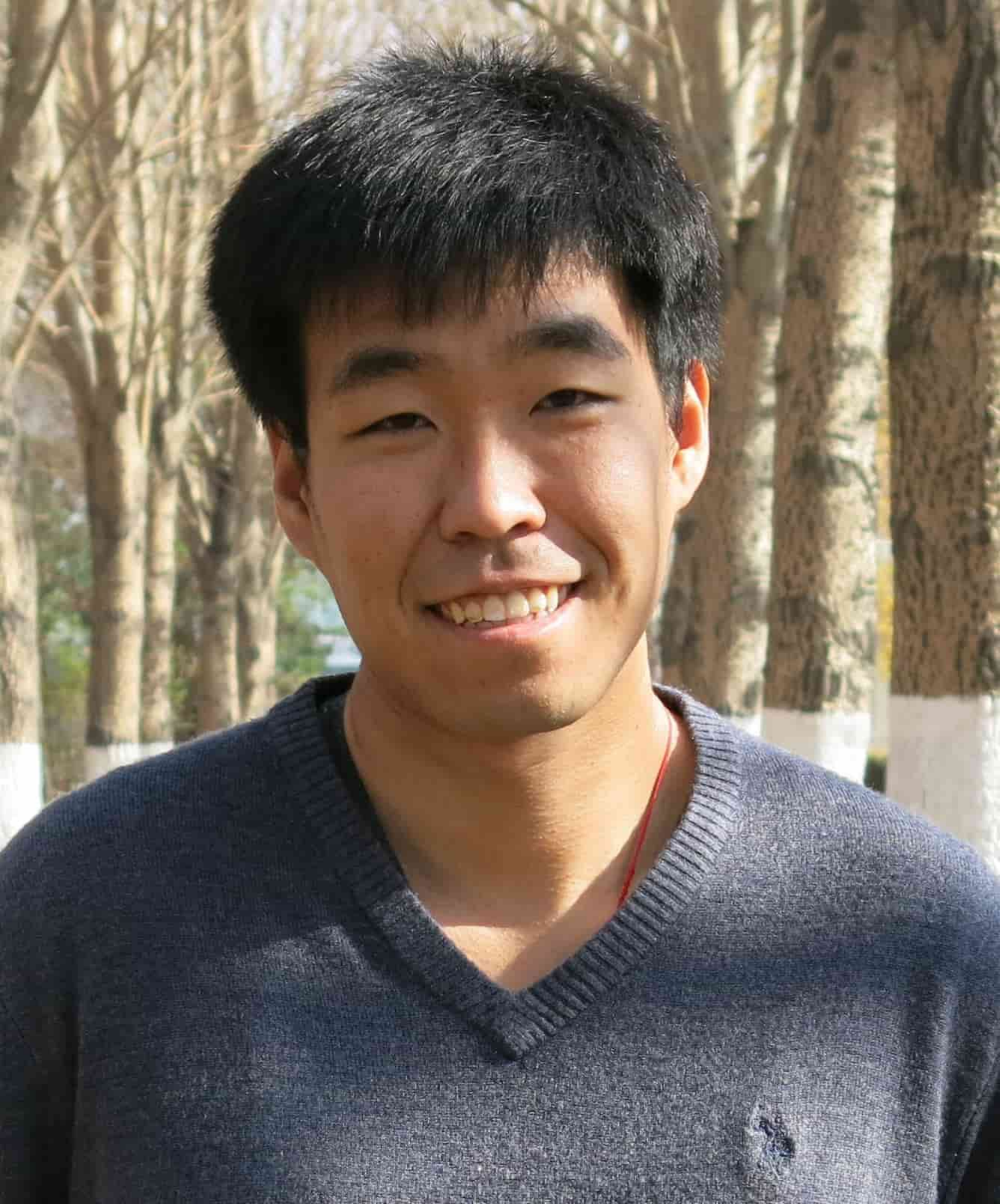}}
	] {Baosen Zhang} received his Bachelor of Applied Science in Engineering Science degree from the University of Toronto in 2008; and his PhD degree in Electrical Engineering and Computer Sciences from University of California, Berkeley in 2013.

	He was a Postdoctoral Scholar at Stanford University, affiliated with the Civil and Environmental Engineering and Management \& Science Engineering. He is currently an Assistant Professor in Electrcial Engineering at the University of Washington, Seattle, WA. His research interests are in power systems and cyberphysical systems. He was selected as one of Forbe's 30 under 30 in energy in 2015.
\end{IEEEbiography}

	\begin{IEEEbiography}
	[
	{\includegraphics[width=1in,height=1.25in,clip,keepaspectratio]{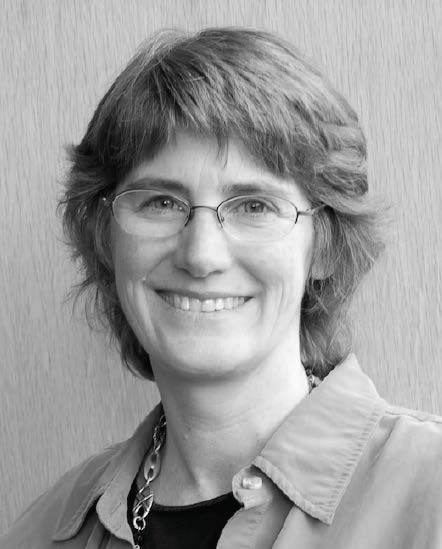}}
	]
{Mari Ostendorf} (M '85, SM '97, F '05) received a PhD in electrical engineering from Stanford University. She has worked at BBN Laboratories, Boston University and is currently an Endowed Professor of System Design Methodologies in Electrical Engineering at the University of Washington. Her research interests are in dynamic, data-driven models for speech and language processing. She is a Fellow of the IEEE and ISCA, a Fulbright Scholar, and winner of the 2010 IEEE HP/Rigas Award and the 2018 IEEE James L. Flanagan Speech and Audio Processing Award.
\end{IEEEbiography}

\begin{IEEEbiography}
	[
	{\includegraphics[width=1in,height=1.25in,clip,keepaspectratio]{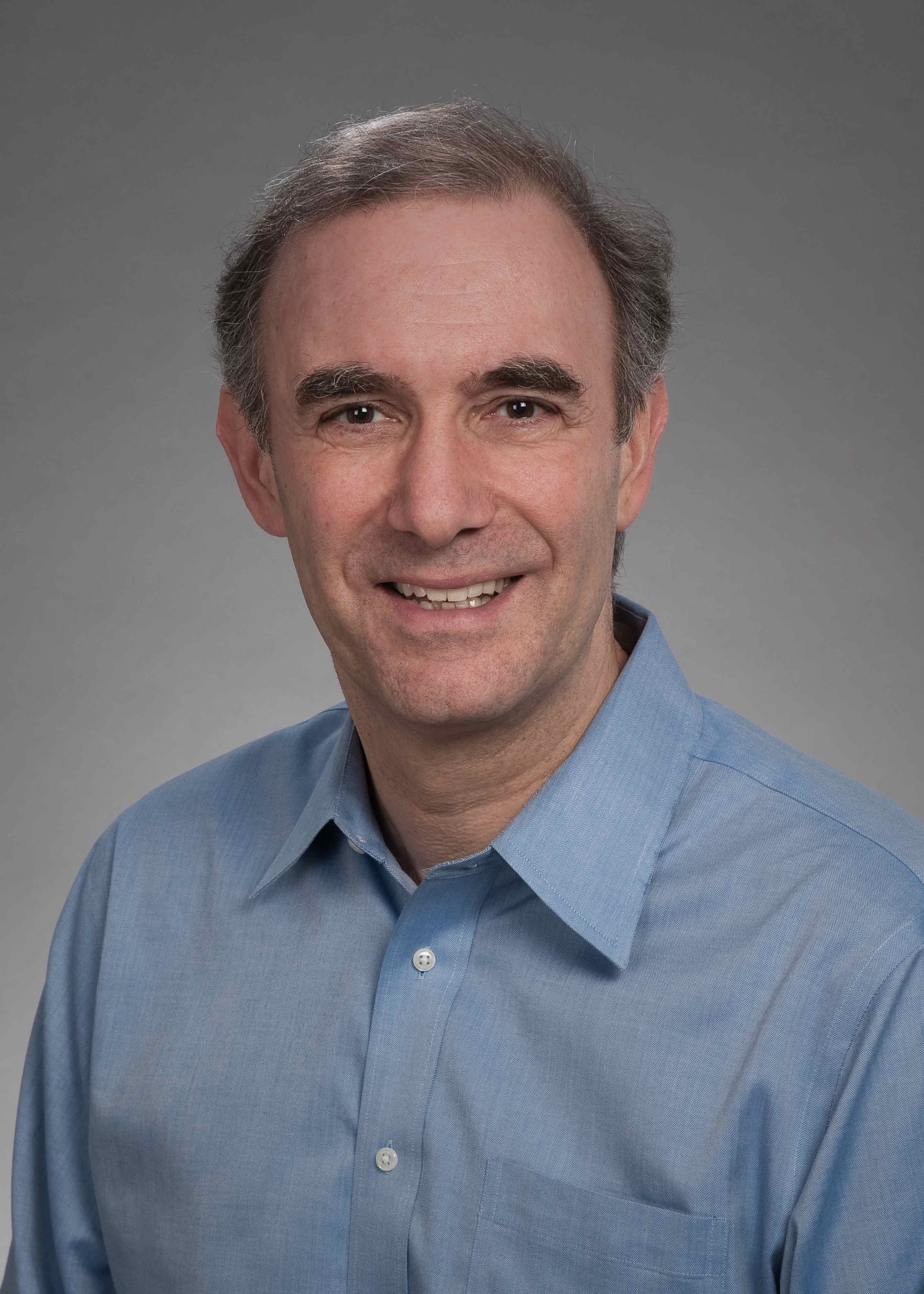}}
	]
  {Daniel Kirschen} is the Donald W. and Ruth Mary Close Professor of Electrical
  Engineering at the University of Washington. His research focuses on the integration of
  renewable energy sources in the grid, power system economics and power system
  resilience. Prior to joining the University of Washington, he taught for 16 years at The
  University of Manchester (UK). Before becoming an academic, he worked for Control
  Data and Siemens on the development of application software for utility control centers.
  He holds a PhD from the University of Wisconsin-Madison and an Electro-Mechanical
  Engineering degree from the Free University of Brussels (Belgium). He is the author of
  two books and a Fellow of the IEEE.
\end{IEEEbiography}

\end{document}

%% file: intro.tex
Outages are fairly common in power distribution networks~\cite{kersting2012distribution,kersting2006recommended}, and this number is increasing in some countries because of aging infrastructure and changing weather patterns~\cite{Caswell2011,Pahwa2007}. While good design and maintenance reduce the number of outages, they cannot be eliminated completely. When an outage is required to perform maintenance or upgrade the equipment, the utility can minimize the disruption of service to  customers by carefully planning the deployment of the crews and the sequence of operations. On the other hand, a fault in the system usually causes an unplanned outages, which can lead to long service interruptions and significant inconvenience to the customers. Therefore, reducing the number of unplanned outages and better managing their duration is a priority for most utilities~\cite{Rustebakke83}. 

The first step towards mitigating the negative consequences of unplanned outages is to gain a better understanding of their number and duration, as well as the number of customers affected. However, by definition, unplanned outages are irregular and difficult to predict~\cite{Kankanala2014}. Most existing studies focus on predicting the \emph{number} of outages and use the weather as the only explanatory variable. Of these, \cite{Zhu2007,Zhou2006,Liu2008,Alvehag2011} attempt to predict the number of outages during extreme weather events~(e.g., hurricanes and ice storms). Other authors \cite{DomijanJrEffectsnormalweather2005,KankanalaEstimationOverheadDistribution2012} try to predict the average number of outages over a given period of time under normal weather conditions. Another line of work aims to rank various components of the power system in terms of their ``susceptibility to failure'' using different machine learning techniques and data sources~\cite{Gross2006,Communications2017,Rudin2012}.

Compared with predicting the number of outages, relatively little work has been done to predict the duration or the total customer-hours lost for a given outage. However, this is arguably the most relevant information from the customers' perspective. When an outage occurs and customers ask when the power will be back on, utilities typically provide an estimate of the restoration time, over the phone, on a website, or using social media. For example, Seattle City Light maintains a real-time outage map with estimated time until restoration.\footnote{http://www.seattle.gov/light/sysstat/map.asp} Since these estimates usually stem from the ``best educated guess'' that operators can produce based on their experience and other factors, the difference between the estimated and the actual outage duration can be quite large~\cite{Chow96}. 

To improve the prediction of outage duration, a number of studies used statistical methods to quantify the relation between various features of outages and their duration. In~\cite{Chow96}, the authors tested the statistical significance of a number of features on the outage duration, but did not provide a specific forecasting algorithm. Adibi and Milanicz ~\cite{Adibi99} developed an estimation method based on the restoration procedure that required a detailed knowledge of the nature of the outage and of the steps that would need to be taken to restore power. This is not practical approach when the goal is to provide customers with an early estimate of how long the outage is like to last because a detailed repair plan is rarely available. Rodriguez and Vargas~\cite{Rodriguez05} designed a fuzzy logic technique that requires a less detailed knowledge of the repair process  but relies on human experts to determine the relative importance of the possible features. This approach combines statistical prediction with some human engineering knowledge, but is somewhat difficult to calibrate and is used as a subroutine in larger restoration optimization problems rather than directly as a reporting tool to the customers.  

In this paper, to predict the duration of outages, \revise{a principled framework  that takes into account exogenous environmental factors (e.g., wind speed and other weather conditions, time of the day), physical features  (e.g., overhead or underground distribution) and engineering knowledge as intrinsically captured in historical outage reports and associated repair logs is proposed.} Information about the ongoing repair process is incorporated incrementally, as it becomes available in the form of entries in the repair log based on reports from the field.

To this end, \revise{a prediction algorithm is trained using a collection of \emph{outage reports} and \emph{repair logs} that most utilities keep, which contain a wealth of information about historical outages.} These records typically include the time and location of the outage, the number of customers affected, its cause, the steps taken to restore service and the outage duration. 
While some data is available in a table,
the repair logs are often written in a ``free writing" style using a combination of colloquial language and very specialized terms. Table~\ref{table:logs} provides an example of such a repair log.\footnote{To protect privacy, addresses, names, and other sensitive information have been replaced by generic labels.} This example shows that these records can be difficult to read even for engineers with domain knowledge. 

\begin{table}[ht]
\caption{Example sequence of outage report and repair logs from an outage.
\label{table:logs}}
\centering
\textbf{Outage Report} \\ \vspace{0.5em}
\begin{tabular}{rc}
\textbf{Outage Start} & 07/21/2009 1:52 pm \\
\textbf{Outage End}   & 07/22/2009 4:30 am \\
\textbf{Feeder}       & 2184               \\
\textbf{Line Type}    & Underground        \\
\textbf{Cause}        & Equipment Failure  \\
\textbf{Customers}    & 113               
\end{tabular} \\
\vspace{1em}
\textbf{Repair Logs} \vspace{1.5em}
\begin{tabular}{cp{7cm}} 
\textbf{Time} & \multicolumn{1}{c}{\textbf{Remarks}} \\ \hline
6:00pm & EMS Urgent alarm and breaker trip. 2184 has cycled. \\
6:03pm & SFD reports bang and brush fire @ 123 Cherry St \\
6:43pm & SLSVC reports N phase lightning arrestor blown @ S/s of S Cherry St, 5 E of 14 Av S. Need URD. \\
6:45pm & NAS reports upon arrival at DU, found 2184 relayed out, is in on auto with 1 Reclosure, targets IOC Phase 1 and Residual \\
7:16pm & J Smith is responding to repair BO lightning arrestor @ S/s of S Cherry St, 5 E of 14 Av S. \\
4:13am & J Smith had 26kv cables, terminators, and electronic sectionalizers on TP 231 to and including cables and BLRs on P-603 to replace BO lightning arrestor. TP 231 @ S/s of S Cherry St, 5 E of 14 Av S. P-603 @ Company XYZ, 123 S Cherry St.
\end{tabular}
\end{table}

To systematically process these logs, \revise{recent advances in the fields of machine learning and \emph{natural language processing} are leveraged to develop an algorithm for real-time outage prediction.}  An initial outage duration (or, repair time) prediction is made based on the environmental factors and physical features available at the start of the outage. As each repair log entry is received, it is summarized using a recurrent neural network (RNN) to provide a vector-space representation that can be easily integrated with physical features for predicting outage duration. Another RNN is used to incrementally update the predicted outage time, incorporating the repair log summary and updating a state vector characterizing the outage. Experiments on a large collection of outage reports demonstrate good performance with the initial predictions and improved results with incremental updates.

This paper is organized as follows. Section~\ref{sec:setup} defines the problem. Section~\ref{sec:model} describes in detail the machine learning methods. Section~\ref{sec:expts} reports on a case study based on Seattle City Light outage data. Section~\ref{sec:analysis} provides examples to illustrate what the model learns from the language of the repair logs. Finally, Section~\ref{sec:conclusion} draws conclusions.



%% file: setup.tex
\begin{figure}[ht]
\centering
\includegraphics[width=0.4\textwidth]{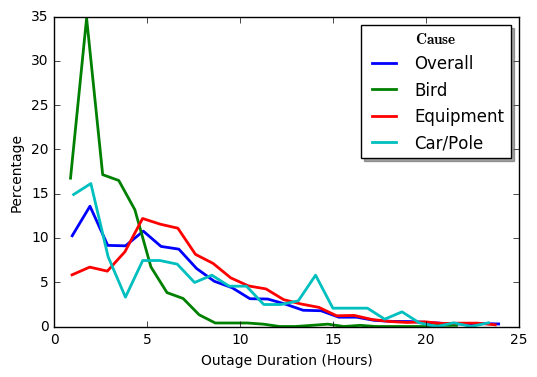}
\caption{Distribution of outage duration for different causes.}
\label{fig:duration_by_cause}
\end{figure}

Utilities measure the impact of an outage in terms of the total customer-hours lost. Using a distribution management system, the number of customers affected by a given outage can be determined fairly easily and accurately~\cite{Rustebakke83,kersting2006recommended,Adibi99}. While this metric is useful for regulatory reporting, from an individual customer's perspective, what matters is the expected duration of the outage. To maintain good customer relations, many utilities try to provide such estimates. Simply providing an average duration calculated over all outages is inaccurate and unhelpful~\cite{Duffey13}. Accurately predicting the duration of an outage is difficult because the repair and restoration process is complex, dynamic and affected by many variables.
\revise{To illustrate this point, Fig.~\ref{fig:duration_by_cause} shows the distribution of outage durations associated with three different causes, based on a collection of 15 years of records of unplanned outages provided by Seattle City Light (SCL) and used in this study.}
As Fig.~\ref{fig:duration_by_cause} shows, even faults with the same cause can lead to very different outage duration.

\begin{figure}
    \centering
    \includegraphics[width=0.45\textwidth]{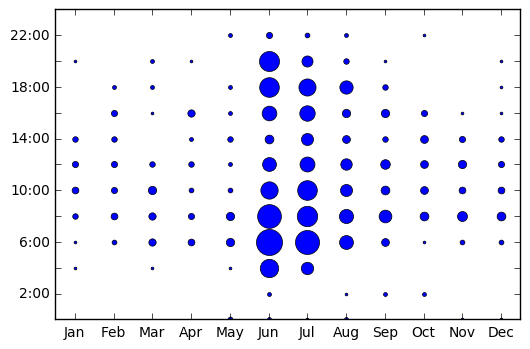}
    \caption{Frequency of outages caused by crows in Seattle. The area of each circle is proportional to the frequency.}
    \label{fig:crow}
\end{figure}

The actual cause of an outage is typically not known when it is first reported, but time of day, season, weather and other factors can provide information that is predictive of the cause.
For example, the SCL outage records show that most damage from crows occurs during the summer in the early morning or late afternoon (See Fig.~\ref{fig:crow}), so knowing that information can lead to good early predictions for bird-related outages.

In addition to giving their customers an expected repair time, utilities may also wish to provide them more detailed probabilistic information, such as the 80\% confidence interval of the outage duration.
\revise{Therefore in this paper a probabilistic forecast of the duration of an outage is provided. Specifically, at a particular time, the proposed method provides a \emph{Gamma} distribution describing the duration using the information collected up to that time.} The Gamma distribution is the probability density function of a non-negative valued random variable, and can be written as
\begin{equation}\label{eqn:gamma}
p(d|k, \theta) = \frac{1}{\Gamma(k)\theta^k}d^{k-1}e^{-\frac{d}{\theta}},
\end{equation}
where $\Gamma(\cdot)$ is the Gamma function, $k$ is the shape parameter and $\theta$ is the scale parameter. By setting $k$ and $\theta$ to different values, the Gamma distribution includes the exponential and Chi-squared distributions, and is commonly used to describe waiting times in many applications~\cite{ramakumar1993engineering,montgomery2009engineering}.


The initial prediction should ideally be refined to take into account new information about an outage as it becomes available from field reports.
\revise{This paper develops and tests such a forecasting approach using the SCL records, which include both repair logs and outage reports, and hourly historical weather information for downtown Seattle.} The logs contain 15 years of records with over 8,000 unplanned outage events and over 40,000 repair logs.\footnote{\revise{Interested readers can request the data from Seattle City Light at {\tt https://data.seattle.gov}}} Since the repair logs are written in ``free-form'' technical English, \revise{a natural language processing tool is needed to combine these inputs with weather information to predict the duration of the outage.}

%

%% file: model.tex
At a high-level, the approach is based on the assumption that the conditional distribution of outage duration $d$ (or repair times) given a set of variables $f$ (e.g. weather, time of day) is reasonably well modeled by a Gamma distribution: $p(d|f)\sim \Gamma (k(f),\theta(f))$. A neural network is used to provide a non-linear mapping from the feature vector $f$ to the parameters $\{ k(f),\theta(f)\}$. The estimated parameters can be used with the Gamma distribution assumption to provide an estimate of outage duration $\hat d = E[D|f]=k(f)\theta(f)$, or they can be used to estimate the 90-th percentile outage duration, for example.

The real-time prediction model provides a sequence of estimates, leveraging multiple neural networks. The initial estimate uses a feedforward neural network to predict the Gamma distribution parameters (Sec.\ \ref{sec:init-est}) using only features available at the onset of the outage (Sec.\ \ref{sec:init-features}). The real-time prediction update model is a Recurrent Neural Network (RNN) that integrates the onset features with a continuous, vector space representation of the incoming repair logs (referred to as an `embedding') and iteratively updates an outage state vector (Sec.\ \ref{sec:RT-est}). The embedding of a repair log is generated using a bi-directional (forward and backward) RNN leveraging an attention mechanism (Sec.\ \ref{sec:repair_embedding}). 

\subsection{Initial Outage Duration Predictor}
\label{sec:init-est}

A feedforward neural network is used to predict a distribution of repair times parameterized by the gamma distribution.
The input to the neural network is $f_i$, the feature vector for the $i$-th outage. The neural network uses two layers with ReLU activations to compute a hidden state vector $g_{i,2}$: 
\begin{equation}
    \begin{split}
        g_{i,1} &= \mathrm{ReLU}(\mathbf{W_1} f_i + b_1) \\
        g_{i,2} &= \mathrm{ReLU}(\mathbf{W_2} g_{i,1} + b_2)
    \end{split}
\end{equation}
The ReLU function, $\mathrm{ReLU}(x) = \max(0, x)$, has been shown to be effective in multi-layer neural networks \cite{glorot2011deep}.

The two parameters of the gamma distribution are directly predicted using $g_{i,2}$.
\begin{equation}
    \label{eq:k_and_theta}
    \begin{split}
        k(f_i) &= \mathrm{softplus}(w_k^T g_{i,2} + b_k) \\
        \theta(f_i) &= \mathrm{softplus}(w_\theta^T g_{i,2} + b_\theta)
    \end{split}
\end{equation}
The $\mathrm{softplus}(x) = \log(1 + \exp x)$ activation function is used on the output layer to ensure that $k,\theta > 0$, as required by the Gamma distribution.

The log likelihood of outage duration $d$, given the features $f$, is computed using $k=k(f_i)$ and $\theta =\theta(f_i)$. \begin{equation}
\log p(d|k, \theta) = \log \Gamma(k) + k \log \theta - (k-1) \log d +\frac{d}{\theta} .
\end{equation}
The objective is to minimize the total negative log-likelihood:
\begin{equation}
-\sum_i \log p(d_i| k(f_i), \theta(f_i)) ,
\label{eq:nll}
\end{equation}
and
the model parameters $\mathbf{W}_1, \mathbf{W}_2, b_1, b_2, w_k, w_\theta, b_k$, and $b_\theta$ are all learned via backpropagation towards that objective.

\subsection{Onset Features}
\label{sec:init-features}

A total of 19 features is available at the onset of the outage. \revise{They are grouped into related categories to help explain their motivation.}

Five features relate to the date and time of the outage: month, day of the week, day of the year, hour of the day and a binary feature indicating if it is a weekend or not. Certain outage types tend to be correlated with the season. Wind and trees are more of a problem in the winter, and bird-related outages are much more frequent during the summer in the early morning or late afternoon. The time of day features and day of week features can also help the model identify the cause of other outages, e.g.\ a car colliding with a pole happens mainly late at night on weekends.

There are nine weather features: temperature, apparent temperature, cloud cover, dew point, humidity, precipitation intensity, precipitation probability, atmospheric pressure, and wind speed. It is likely that not all of these are useful but there is no harm in including them, since \revise{regularization techniques are used to avoid overfitting (see Sec.~\ref{sec:implement})}.

\revise{Two features are used to indicate the difficulty of repairing outages at each location.} The first is a binary feature indicating if the distribution is overhead or underground. The second is a smoothed average of historical repair times for outages from that feeder, where smoothing is a weighted combination of the average for that feeder and the average for all feeders depending on the number of outages observed for that feeder. The last three features provide information about the size of the outage and the busyness of the repair crews. They are the logarithm of the number of customers affected by the outage, the total number of outages in the last three hours, and the total number in the last eight hours.

These features are selected because they are all immediately available at the start of the outage, and have been used in prior work as discussed earlier. \revise{An \emph{oracle} feature is also experimented with in this paper, i.e. a feature that is generally not known until the repair is underway, in this case the cause of the outage.}  Optionally including this feature allows us to test how well our other features implicitly capture the cause.

\subsection{Real-time predictions with repair logs}
\label{sec:RT-est}

\begin{figure}
    \centering
    \includegraphics[width=0.5\textwidth]{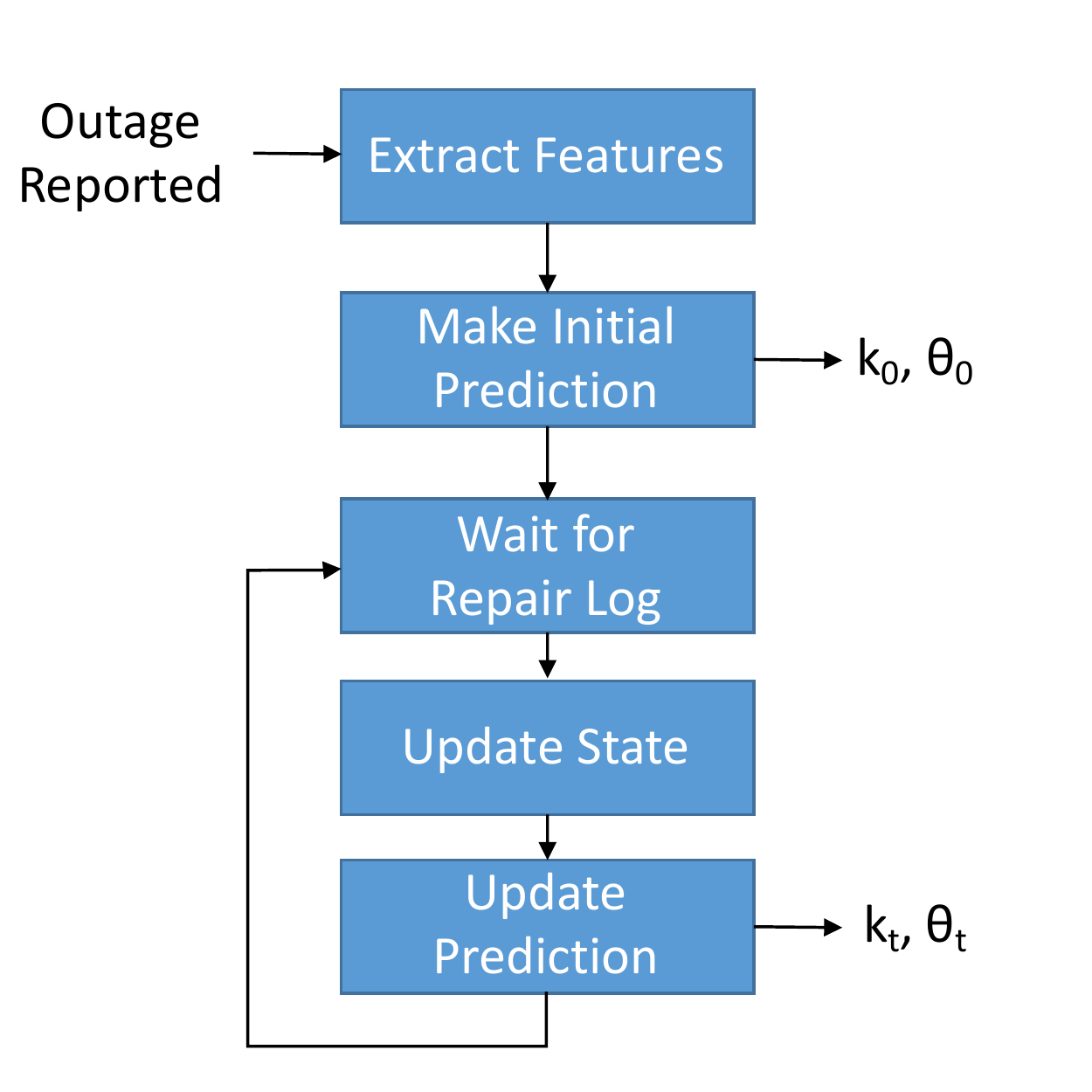}
    \caption{\revise{Flowchart of the real-time prediction system. When the outage is first reported, an initial prediction is given using onset features such as weather, underground/overhead lines, ect. Then as the repair logs arrive, the state of the machine learning algorithm is updated, leading to an update in the predicted outage times.}}
    \label{fig:flowchart}
\end{figure}

Our real-time prediction model makes use of the repair logs to update its predictions during the outage. As described above, an initial prediction is made at the start of the outage. Thereafter, each time a repair log is received, the system extracts relevant information about the progress of the repair and issues a new prediction. \revise{This procedure is depicted in the flowchart in Fig. \ref{fig:flowchart}.}

The updated predictions are driven by a recurrent neural network. At each time step, the RNN takes two vector inputs and produces a vector output $o_t$
\begin{equation}
\label{eq:gru}
o_t = \mathrm{RNN}_U(o_{t-1}, [f, s_t, \log(T + 1)]).
\end{equation}
The first input is the output from the RNN at the previous time step, $o_{t-1}$, which functions as a summary of the state of the outage up to that point. The second is a vector that concatenates the onset features $f$ (Section \ref{sec:init-features}) with a vector summary of the latest repair log $s_t$, plus an additional log-transformed feature indicating the amount of time elapsed since the beginning of the outage, $T$. The method of creating the repair log embedding vector is described in Section \ref{sec:repair_embedding}. 

To compute $o_1$ for the first repair log, \revise{the input $o_0 = \mathbf{P} f$, a projection of the onset features, is used. The matrix $\mathbf{P}$ is learned jointly with the other  $\mathrm{RNN}_U$ parameters.}

The output from $\mathrm{RNN}_U$ is used in \eqref{eq:k_and_theta2} to predict the Gamma distribution parameters 
\begin{equation}
    \label{eq:k_and_theta2}
    \begin{split}
        k_t &= \mathrm{softplus}(v_k^T o_t + \beta_k) \\
        \theta_t &= \mathrm{softplus}(v_\theta^T o_t + \beta_\theta)
    \end{split}
\end{equation}
much like in \eqref{eq:k_and_theta}.

\begin{figure}[ht]
    \centering
    \includegraphics[width=0.48\textwidth]{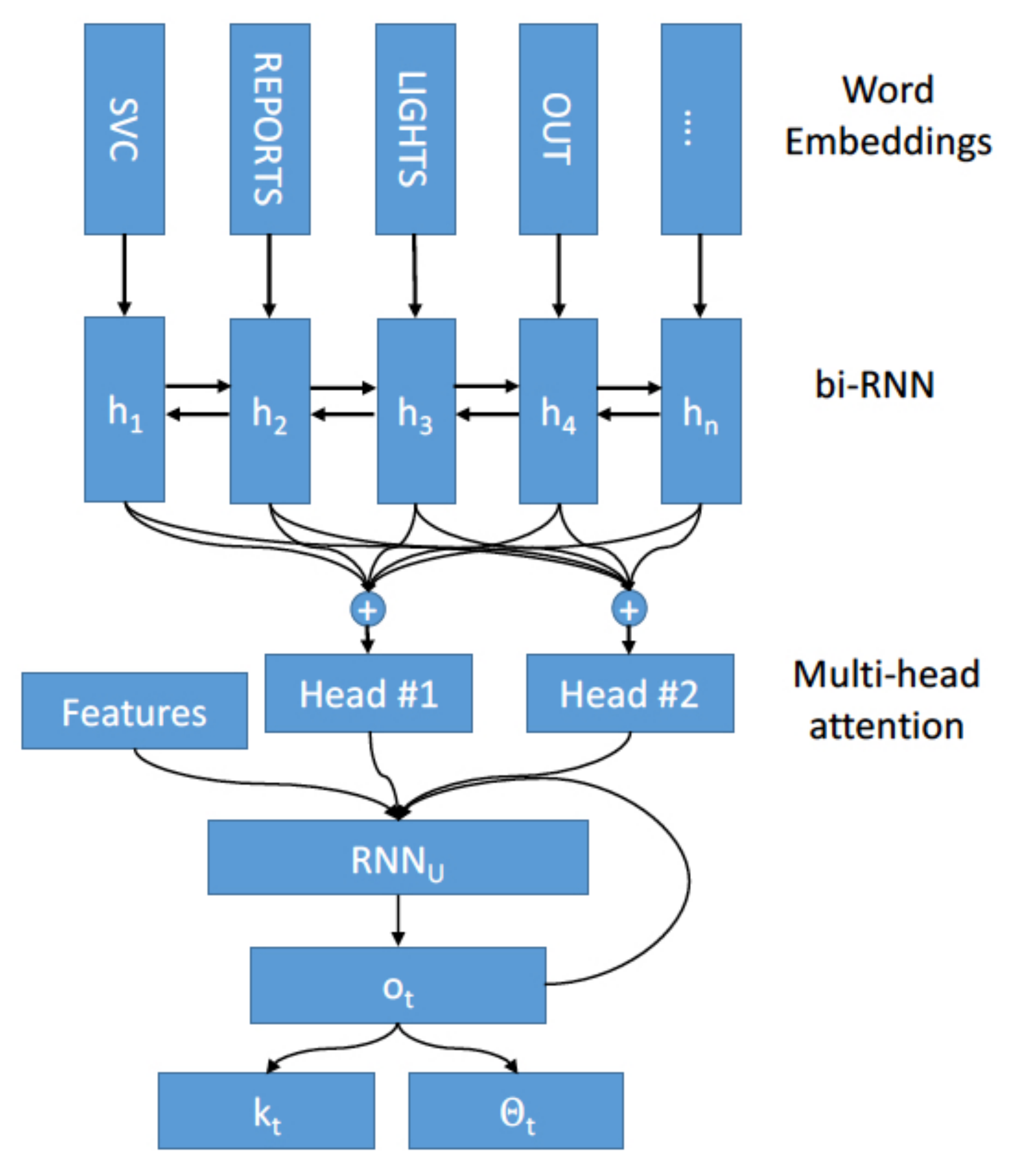}
    \caption{Diagram of the neural network architecture showing how the repair log is used to update the duration prediction. The output of the network are the parameters of the Gamma function that models the distribution of the outage duration.}
    \label{fig:architecture}
\end{figure}

Fig.~\ref{fig:architecture} shows a diagram of the neural network architecture. Again, the model parameters are trained using the negative log-likelihood objective (see \eqref{eq:nll}).

\subsection{Repair Log Embedding Vector}
\label{sec:repair_embedding}

To create the $s_t$ vector that summarizes the $t$-th repair log, we leverage techniques from natural language processing for creating embeddings of short text-like sentences or paragraphs, namely bi-directional RNN's with attention. The bi-directional RNN builds representations that capture the meaning of each word in its local context. Attention is a method of collapsing the per-word representations from the bi-RNN into a single summary vector by taking a weighted combination. 

A vocabulary $V$ is defined by selecting all words\footnote{`Words' are unique white-space-separated tokens, after some preprocessing, described in Sec.~\ref{sec:implement}.} with counts greater than some threshold and adding an out-of-vocabulary token for other words.
Having defined the vocabulary, each word is mapped to a $|V|$-dimensional indicator vector with a single 1 and all remaining elements equal to zero (referred to as a one-hot vector).
Thus, the input sequence of words in the repair log is represented as a sequence of one-hot encoded vectors $w_i$:
\begin{equation}
    S = (w_1, w_2, w_3,\ldots,w_n) .
\end{equation}
These vectors are projected to a low-dimensional embedding space using a matrix $\mathbf{E} \in \mathbb{R}^{k \times |V|}$, resulting in the sequence 
\begin{equation}
    S' = (\mathbf{E}w_1, \mathbf{E}w_2,\ldots,\mathbf{E}w_n) .
\end{equation}

The sequence $S'$ is input to two RNNs: one that processes the input from left to right and another that processes the input from right to left. The combination of these two RNNs is referred to as a bi-directional RNN.
The outputs at each position from each direction of the RNN are concatenated together to create a representation $h_i$ for each word in the log:
\begin{align}
    \label{eq:birnn}
    h_i &= [h^{forward}_i, h^{backward}_i] .
\end{align}
The sequence of all the $h_i$ vectors across the repair log forms a matrix $\mathbf{H}= (h_1, h_2,\ldots,h_n) \in \mathbb{R}^{n \times 2c}$.

Neural attention is a widely used technique that allows the model to summarize a sequence using a weighted average, where the weights are predicted by the model to focus on (or attend to) pieces of information that it judges to be relevant \cite{Bahdanau2014NeuralMT,Santos2016AttentivePN}. We apply attention to the recurrent state vectors $h_i$ in $\mathbf{H}$ to summarize the repair log in a vector $s$. The weights in the average are computed using parameter matrices $\mathbf{M}_1 \in \mathbb{R}^{d \times 2c}$ and $\mathbf{M}_2 \in \mathbb{R}^{2c \times 2c}$ and the $\mathrm{softmax}$ function to create a normalized distribution.
\begin{align}
    q_t &= \mathrm{ReLU}(o_{t-1}\mathbf{M_1} + b_1) \\
    \mathbf{Y}_t &= \mathrm{ReLU}(\mathbf{H}_t \mathbf{M}_2  + b_2) \\
    s_t &= \mathrm{softmax}(q_t\mathbf{Y}_t) \mathbf{H}_t 
\end{align}

The above process can be duplicated with parallel computation of different sets of attention weights using the same $\mathbf{H}_t$ but different sets of parameters $\{ \mathbf{M_1}, \mathbf{M_2},  b_1, b_2 \}$. The resulting vectors are then concatenated to form a high dimension $S_t$. This is referred to as multi-head attention.
Our best performing models use two-headed attention \cite{vaswani2017attention}. 
The motivation for multi-head attention is that it allows each attention head to have a specialized purpose. For instance, one might focus on the cause of the outage and the other might focus on which team will be responding.

%% file: expts.tex
\subsection{Data} \label{sec:data}

\revise{Three sources of data are available: outage reports, repair logs, and weather information.} The outage reports were provided by Seattle City Light and span 15 years of data. The repair logs cover the same period and contain more than 30,000 textual records. (Refer to Table \ref{table:logs} for examples.)

\revise{The outages are divided into training, validation, and test sets based on the date of the outage.} Outages occurring before March 15, 2014 are assigned to training, those between March 15, 2014 and March 15, 2015 are used for validation, and those after March 15, 2015 are for testing. Outages lasting more than 24 hours (which tend to be associated with major storms), planned outages and outages lasting less than 5 minutes (for which predictions are not needed) are not included in our data. There are 6,172 outages in the training set, 740 in the validation set, and 851 in the test set.

The outage data is supplemented with hourly historical weather information.\footnote{Weather data from darksky.net} The weather information is provided for a single location: downtown Seattle. We align the weather data with the outages by selecting the information from the weather report which is closest in time to the start of the outage.

Since the outage reports and repair logs were collected over a period of several years, without anticipating the use of language processing, some work is required to format and clean the data. The repair logs are aligned with the outages by selecting the log entries that were made between the start and end times of the outage on the same feeder. The alignment is not exact as there may be more than one outage at the same feeder at the same time, but such cases are rare. Approximately 20\% of the outage events do not align with any repair logs for a variety of reasons, e.g. transmission-level outages which are handled through a separate process. \revise{These outages are not included in the experiments that make use of repair logs.} In some cases, a repair log will be made to note the conclusion of the outage. \revise{Logs that occur in the last 2.5\% of an outage duration are removed} because it is not useful to make further predictions at that point and these logs interfere with the fit of the model. \revise{For the real-time predictions there are 19,182 repair logs in the training data, 2,403 in the validation set and 3,155 in the test data.}

\subsection{Implementation Details}
\label{sec:implement}

The model is implemented using the Tensorflow library \cite{tensorflow2015}. Fitting is done using the Adam optimizer with a learning rate of $0.001$  and a batch size of one \cite{kingma2014adam}. All of our recurrent neural networks are of the Gated Recurrent Unit \cite{cho2014learning} variety with layer normalization \cite{ba2016layer}. There are two regularization strategies: early stopping \cite{caruana2001overfitting} and variational dropout on the GRUs \cite{kingma2015variational}. Model code is available on GitHub.\footnote{http://gitub.com/ajaech/outageduration}

Punctuation is removed using a simple regular expression. The repair log text is preprocessed by lower-casing and by replacing ID numbers with their types such as for transformers, feeders, or poles. (An example is found in Fig.~\ref{fig:heatmap2} where the telephone pole identifier is replaced with \textless tp\textgreater.) The vocabulary is set by taking all words that appear in the training data more than a certain number of times where the cutoff is selected during tuning. The vocabulary size of the best models ranges from two thousand to four thousand words.

Hyperparameter tuning is done using a random search strategy, selecting the model that assigns the highest likelihood to the validation data. The hyperparameters are the vocabulary cutoff, the word embedding size, the $\mathrm{RNN}_U$ GRU cell size, the bi-directional GRU cell size, the dropout rate, the number of epochs to train, number of attention heads (one or two), and whether or not to use layer normalization. We find that early stopping is a better regularizer than variational dropout, layer normalization is helpful, and two attention heads is better than one. 

\subsection{Initial Outage Duration Prediction Results}
The metrics are negative log likelihood (the training objective), root mean squared error (RMSE), and Pearson's correlation. The negative log likelihood is a measure of both how well the model is able to predict the true duration and also how well it is able to reduce the uncertainty of its predictions. Since the model is trained based on a negative log likelihood objective, improvements to the model are best observed with this measure, but it is less interpretable from an applications perspective. For comparison, a linear regression model was trained to optimize for mean squared error. The linear regression gives no uncertainty information and is slightly worse in terms of RMSE (4.3 hours) and correlation (28.7) for the all onset features condition. Results for other feature sets with this model are similar. 

Table \ref{table:main_results} presents the experimental results. The case with no features corresponds to using a single gamma distribution for all outages. As more features are added, the model achieves a better negative log likelihood, i.e.\ provides a better fit of the observed test data, lower RMSE, and higher correlation. The last two lines serve as oracle experiments, since they include the true cause of the outage as a feature, which is not usually known at the onset. As expected, knowing the true cause improves performance for all metrics. We hypothesized that the onset features would give us some information about the true cause, which seems to be the case.
A classifier trained to predict the outage cause from the onset features has an accuracy of 70\%. (Always predicting the majority class `Equipment Failure' gives an accuracy of 44\%.)
\begin{table}[ht]
\centering
\caption{Performance of initial outage duration predictions with different feature sets, as measured using negative log likelihood (NLL), root mean squared error in hours (RMSE), and Pearson's correlation.}
\label{table:main_results}
\begin{tabular}{crrr}
\textbf{Feature Set}   & \multicolumn{1}{c}{\textbf{NLL}} & \multicolumn{1}{c}{\textbf{RMSE}} & \multicolumn{1}{c}{\textbf{Corr.}} \\ \hline
No Features            & 2.72                             & 4.45                              & 0.0  \\
Weather                & 2.71                             & 4.41                              & 12.1                                \\
Time                   & 2.71                             & 4.41                              & 16.3                               \\
Time + Weather         & 2.70                             & 4.40                              & 18.0                               \\
All Onset Features     & 2.66                             & 4.25                              & 32.0                               \\ \hline
Cause Only             & 2.60                             & 4.15                              & 36.9                               \\
Cause + Onset Features & 2.57                             & 4.02                              & 45.1                              
\end{tabular}
\end{table}

Using a gradient boosted regression tree to assess feature importance, the top five features are the average outage duration for that feeder, the customers affected, the hour of the day, the day of the year, and the air pressure. The day of year feature is helpful because of the seasonality of different outage types. When binary indicator oracle features are added for each outage cause category then the feature importances are similar except that the Bird/Animal cause indicator is ranked as the sixth most important feature.

\subsection{Real-time Prediction Results}

\begin{figure}
    \centering
    \includegraphics[width=0.33\textwidth]{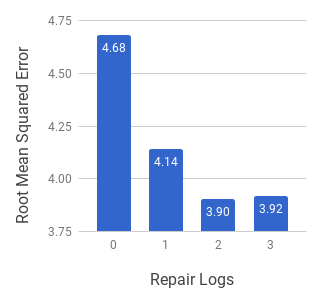}
    \includegraphics[width=0.33\textwidth]{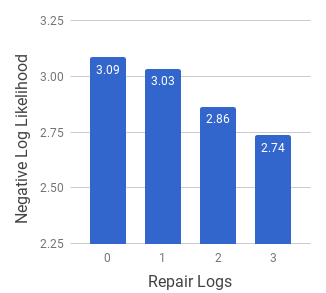}
    \caption{Improvement in negative log likelihood and root mean squared error in the real-time prediction system.}
    \label{fig:barcharts}
\end{figure}

Figure~\ref{fig:barcharts} demonstrates the performance of the real-time prediction system by showing the root mean squared error and the negative log likelihood for the initial prediction and after receiving one to three repair logs. These metrics are computed on the subset of the test data where there were at least three repair logs for each outage. Because of this, the results for the real-time prediction are not comparable with our previous experiments. Both metrics show a trend of increased prediction accuracy as more information from the repair logs becomes available.

%% file: analysis.tex
\begin{figure}
    \centering
    \includegraphics[width=0.48\textwidth]{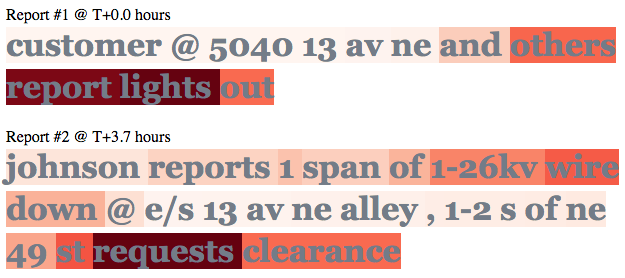}
    \caption{Heat map of attention for a selected outage. Darker colors received greater weight. In the first report the attention is given to words stating that fact that power is out. In the second report the attention focuses on words identifying the cause (wire down) and possible actions (requests clearance).}
    \label{fig:heatmap1}
\end{figure}
\begin{figure}
    \centering
    \includegraphics[width=0.4\textwidth]{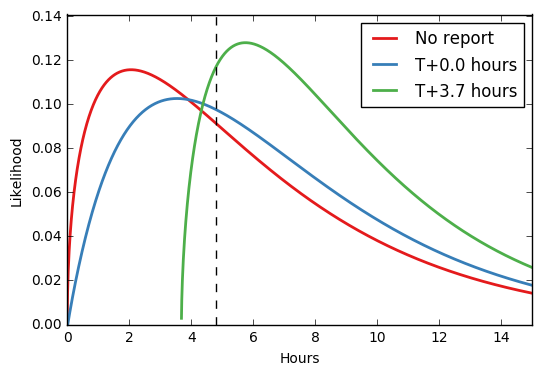}
    \caption{Predicted outage duration distributions for the same 4.8 hour outage as Fig.~\ref{fig:heatmap1}. The dotted line marks the true outage duration and the most likely durations predicted by each distribution are 2.1, 3.5, and 5.7 hours respectively. As more reports come in, the predictions successively improve.}
    \label{fig:predicted_distribution1}
\end{figure}

Figures \ref{fig:heatmap1} and \ref{fig:predicted_distribution1} illustrate an example where successive repair log progressively improves the prediction. In Fig.~\ref{fig:heatmap1}, a heat map visualization shows where in the the repair logs the model is placing its attention. Since the vector associated with a word in the text is a concatenation of the forward and backward stages of the bi-RNN, it encodes information from the surrounding phrase. To make this more clear in the heat maps, the word attention weights are smoothed over the sequence. Observe that the model identifies the cause ``1-26kv wire down'' and the phrase ``requests clearance'' which tends to be associated with a speedy repair from that time point. Figure  \ref{fig:predicted_distribution1} shows the predicted distributions of outage duration evolve as field reports are received, as well as the actual outage duration. The distributions correspond to time remaining, so their start time is the arrival time of the corresponding report. The ``no report'' condition uses only the onset features.


\begin{figure}
    \centering
    \includegraphics[width=0.48\textwidth]{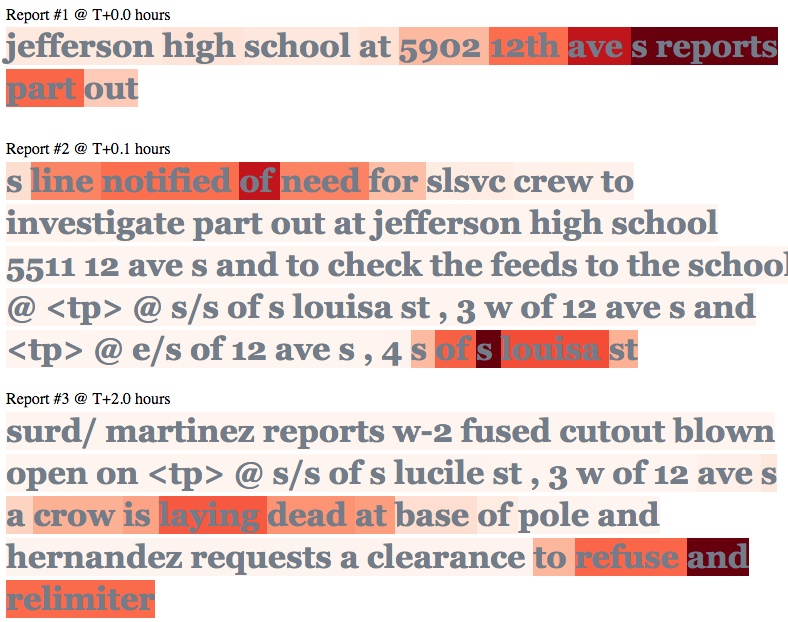}
    \caption{Heat map of attention for an outage where some of the reports do not help with prediction (see Fig.~\ref{fig:predicted_dist2}). In each of the log entries, darker colors received greater weight. As we can see, in the top entry, there is no useful information to put the weights on except the fact that a line is out. By the next entry in the log, the algorithm is able to pick out the cause of the outage (dead crow) and the action (refuse and relimiter).}
    \label{fig:heatmap2}
\end{figure}

\begin{figure}
    \centering
    \includegraphics[width=0.4\textwidth]{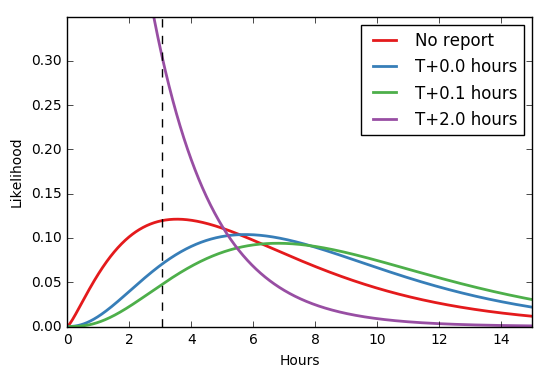}
    \caption{Predicted distributions of outage duration for a 3.0 hour outage (reports in Fig.~\ref{fig:heatmap2}) where the 1st and 2nd report did not help increase prediction accuracy. After these reports, the distribution actually shifted away from the true report time because the cause was not clearly identified. After the 3rd report correctly identifies the cause as a dead crow, the outage duration prediction improves dramatically. \revise{(The final distribution is an exponential that starts at T=2.0; the full height is not illustrated.)}  The most likely durations predicted by each distribution are 3.5, 5.7, 6.7, and 2.0 hours respectively.}
    \label{fig:predicted_dist2}
\end{figure}

Figures \ref{fig:heatmap2}
and \ref{fig:predicted_dist2} illustrate a case where the first and second reports do not improve the prediction. They indicate uncertainty on the cause of the outage, which increases the expected duration. Because the cause (i.e. dead crow) is identified in the 3rd report, a speedy repair can then be predicted. The probability distribution for the final prediction extends beyond the top of the figure and is truncated to improve readability.

\revise{
The distributions predicted could be used in a variety of ways to update customers on the status of repairs. For example, the time could be adjusted to be more or less conservative; e.g., Table~\ref{table:example} shows mode, mean (min MSE), and 80\% confidence estimates of time remaining until power is restored for the 3-report example above. In addition, the attention weights could be used to report a cause when it is reliably identified, such as the bird-related fault identified in report 3. The specific strategy used should be assessed with customer studies.
}

\begin{table}
\centering
\caption{Example of options for reporting time remaining until power is restored for distributions predicted in Fig.~\ref{fig:predicted_dist2}.
\label{table:example}}
\begin{tabular}{|l|c|c|c|c|}\hline
Report Time & True & Mode & Mean & 80\% \\ \hline
Onset & 3.0 & 3.5 & 6.2 & 9.2 \\
$t=0.01$ & 3.0 & 5.7 & 5.1 & 11.4\\
$t=0.1$ & 2.9 & 6.7 & 6.3 & 12.9\\
$t=2.0$ & 1.0 & 0.0 & 2.0 & 3.1\\ \hline
\end{tabular}
\end{table}


\begin{table}[]
\caption{Top bigrams from the repair logs attended to by each of the attention heads.}
\label{table:bigrams}
\centering
\begin{tabular}{cr|cr}
\multicolumn{2}{c}{\textbf{Head \#1}}                 & \multicolumn{2}{c}{\textbf{Head \#2}}                                              \\
\textbf{Bigram} & \multicolumn{1}{c}{\textbf{Count}} & \textbf{Bigram}                              & \multicolumn{1}{c}{\textbf{Count}} \\ \hline
26kv cables      & 185                                & \textless CL\textgreater \textless END\textgreater& 209                                \\
duty supervisor  & 139                                & to investigate                                & 120                                \\
26kv line        & 109                                & lights out                                    & 73                                 \\
need nurd        & 92                                 & need nurd                                     & 47                                 \\
lights out       & 84                                 & to respond                                    & 44                                 \\
part out         & 51                                 & slsvc to                                        & 39
\end{tabular}
\end{table}

It is informative to analyze the most common phrases attended to by the two attention heads. Table \ref{table:bigrams} summarizes the most frequent bigrams (adjacent word pairs) for each head.
To create this table, we find the word $w_i$ in each report that is given the highest weight and count the two bigrams associated with that word: $(w_{i-1},w_i)$ and $(w_e,w_{i+1})$. The two heads specialize in different concepts but there is some overlap. The first head frequently identifies the mention of 26kV cables. The second head frequently identifies the inclusion of the term \textless CL\textgreater, a marker for an ID number of a report that is typically created in the log towards the end of an outage.

%% file: concl.tex
This paper introduces an approach for predicting outage duration by learning from historical outage records. It also shows how natural language processing can be used to provide additional features allowing real-time updates of duration estimates. Experiments with a large collection of outages show that good results can be obtained from environmental features alone, since there is good correlation between these features and some causes. In addition, improvements are possible by using text analysis of incoming repair logs that provide information related to the outage cause and repair steps. 

The model proposed here was developed to predict a distribution of duration times, from which one could predict either the expected time until service is restored or a time within which there is a certain level of confidence that service will be restored. The framework could just as easily be used to predict the estimated time to repair directly by replacing the final neural network layer ($k$, $\theta$ estimators) with an expected duration prediction layer, and changing the training objective to mean squared error. \revise{Experiments found that the RMSE results are only slightly better when optimizing directly for that objective than when using the gamma distribution.}

The model proposed here advances on prior work but is also complementary. For example, in \cite{Kankanala2014}, it is shown that an ensemble of neural networks is an effective strategy for predicting the number of outages specifically looking at weather related (wind and lightning) factors. 
In contrast, the work here addressed prediction of outage duration and considered all types of outages (excluding major storms), but the benefits of ensembling may extend to outage prediction. 
Work on outage duration prediction that relies on environmental factors (vs.\ post-hoc knowledge of the cause) has investigated the importance of different factors \cite{Chow96}, which motivated many factors explored here. However, prior work did not integrate these in a unified model. Because different factors interact (e.g.\ time of day and season for bird-related outages), it is useful to explore integrated models.

\revise{A constraint of the approach described here is that it requires historical distribution data associated with the region covered. As learned in this study, weather patterns impact the prediction, as does the type of infrastructure. If such historical data was available from a several cities, it would be possible for the initial prediction model to learn to generalize to a new urban area. The text-based updates will be more sensitive to the idiosyncrasies of reporting in a particular region.}

\revise{There are a number of opportunities that the use of historical records and natural language processing could enable in future studies. For example, the data could be used to predict the likelihood of failure for particular types of equipment in the next few years.  Further analysis of the attended words could provide guidance as to what sort of information should be included in field reports and provide automated suggestions about outage causes and repairs to engineers in the field.} 